# Tunable, grating-gated, graphene-on-polyimide terahertz modulators

Alessandra Di Gaspare[1], Eva A.A. Pogna[1], Luca Salemi[1] Osman Balci[2], Alisson R. Cadore[2], Sachin M. Shinde[2], Lianhe Li,[3] Cinzia di Franco[4], A. Giles Davies,[3] Edmund Linfield,[3] Andrea C. Ferrari[2], Gaetano Scamarcio[4] and Miriam S. Vitiello[1]

[1]NEST, CNR-NANO and Scuola Normale Superiore, 56127 Pisa, Italy
[2]Cambridge Graphene Centre, Cambridge CB3 0FA, UK
[3]School of Electronic and Electrical Engineering, University of Leeds, Leeds LS2 9JT, UK
[4]CNR-IFN and Dipartimento Interateneo di Fisica, Università degli Studi di Bari, I-70126 Bari, Italy

We present an electrically switchable graphene terahertz (THz) modulator with a tunable-by-design optical bandwidth and we exploit it to compensate the cavity dispersion of a quantum cascade laser (QCL). Electrostatic gating is achieved by a metal-grating used as a gate electrode, with an $HfO_2/AlO_x$ gate dielectric on top. This is patterned on a polyimide layer, which acts as a quarter wave resonance cavity, coupled with an Au reflector underneath. We get 90% modulation depth of the intensity, combined with a 20 kHz electrical bandwidth in the 1.9–2.7 THz range. We then integrate our modulator with a multimode THz QCL. By adjusting the modulator operational bandwidth, we demonstrate that the graphene modulator can partially compensates the QCL cavity dispersion, resulting in an integrated laser behaving as a stable frequency comb over 35% of the laser operational range, with 98 equidistant optical modes and with a spectral coverage of ~ 1.2 THz. This has significant potential for frontier applications in the terahertz, as tunable transformation-optics devices, active photonic components, adaptive and quantum optics, and as a metrological tool for spectroscopy at THz frequencies.

## Introduction

Key photonic applications in the far-infrared, i.e. 1–10 THz frequencies, require modulation and switching of the optical signals with high speeds (< 10 kHz), large (> 50%) amplitude modulation depths and frequency tuning [1]. E.g. high-throughput wireless transmission over short-range links requires an efficient (> 50%) intensity modulation [2]; fast (< $\mu$s) spatial light modulation is desirable for high-resolution imaging [3] and spectroscopy systems; terahertz (THz) communications rely on the development of fast reconfigurable components for amplitude, frequency and phase stabilization [4, 5]. The terahertz range (0.1-10 THz) is interesting for future high speed communications since the high carrier frequencies would allow unprecedented channel capacities. As an example, Ref. 2 reported single-input and single-output wireless communication systems at 237.5 GHz for transmitting data over 20 m, at a data rate of 100 Gbit s$^{-1}$.

Miniaturized quantum cascade laser (QCL) sources, operating at THz frequencies [6,7] can support very high modulation rates (up to tens of GHz) [8,9] through direct modulation of their operating current [10,11]. However, they require cryogenic [7] or Peltier cooling [12-14]. Most importantly, the modulation of their driving current often induces undesired effects, such as current instabilities [6,7], or a simultaneous amplitude and frequency self-modulation [15]. Hence the need to use small cavities, with consequently low ($\mu$W) power outputs that give major drawbacks for applications. Thus, electro-optical modulators independent from source-detector combination, or integrated with a specific source, are highly desirable.

Single layer graphene (SLG) with its broadband and electrically tunable optical conductivity [16] is an ideal platform for the development of electrically switchable electro-optic modulators [17-19]. The optical conductivity of SLG is defined by the interband and intraband electronic transitions between or within the conduction band (CB) and valance band (VB) [17,20]. In the visible, near-infrared and mid-infrared, interband transitions dominate the optical conductivity of SLG [20, 21]. In intrinsic SLG, the light absorption in IR and visible is $\sim \pi\alpha \cong 0.023$, normalized to the incident light intensity, where $\alpha$ is the fine structure constant [21]. By shifting the chemical potential below or above the half frequency of impinging radiation via electrostatic gating, the absorption can be tuned, owing to Pauli blocking [19]. High modulation amplitudes can then be achieved by employing multiple SLGs [17, 22]. At THz frequencies, however, interband transitions are usually blocked due to the fact that as-prepared SLG is doped [23], and due to temperature fluctuations [24], hence intraband electronic transitions dominate the optical conductivity [25]. Consequently, SLG behaves as a two-dimensional electron gas and has a Drude-like conductivity in the form of $\sigma = iD/\pi(\omega + i\Gamma)$, where $\omega$ is the light frequency, $\Gamma$ is the scattering rate, $D = (v_F e^2/\hbar)\sqrt{\pi|n|}$ with $v_F$ the Fermi velocity, $n$ the charge carrier density [26, 27, 28]. This can be controlled by means of either electrostatic gating [28] or optical excitation [25] over a spectral range significantly broader compared to conventional semiconductors [29]. Tuning $\sigma$ enables control of the light-SLG interactions by means of transmission, reflection and absorption [17,19]. For free standing SLG, higher $\sigma$ results in higher reflection and lower transmission of THz-frequency light, and the absorption reaches its maximum when the sheet resistance ($R_s$) of SLG reaches half of the free space impedance ($Z_0 = 377\ \Omega$)[30].

The progress in the large area growth and transfer of SLG [23, 31], recently enabled the development of graphene-based active devices such as modulators [18], absorbers [32], phase shifters [5], and reflect-arrays [33] at THz frequencies [34-36]. In all these devices, phase and amplitude control of the incident light were achieved by tuning the SLG charge density by all-electronic [37, 38] or all optical [39, 40] architectures. A variety of configurations were used to design SLG-THz reflection modulators. These includes: architectures based on placed on $SiO_2$/undoped Si, with an Au reflector on the back and a ring-shaped electrode on the SLG [37], in which the charge density of SLG was tuned via the back-gate, achieving amplitude modulations between 15% and 64% at 20 kHz speed between 0.59 and 0.63 THz [37]; modulators relying on plasmonic metamaterials, exhibiting narrow (< 100 GHz) bandwidths, large modulations (> 50%) enabled by the field enhancement, and switching speeds > 10 MHz in the sub-THz range [41]; configurations exploiting the change in conductivity of multilayer graphene (MLG) to externally modulate change in conductance of a graphene multilayer to modulate the LC plasmonic resonance of a periodic pattern of metallic meta-atoms externally [42], achieving 58% amplitude modulation, phase modulation of 65° and a 12 MHz speed at 0.8 THz; split-ring resonators strongly coupled to graphene surface plasmons [43], allowing 60% modulation at a fixed and very sharp frequency (4.7 THz) with 40 MHz speeds; and, Brewster angle devices [44] and chiral metamaterials [45,46], all operating at frequencies < 1.5 THz, amongst many other schemes [47,48].

Although the field is very vibrant, the combination of high modulation efficiencies (> 60%) and modulation speeds < 100 μs, with tunable and broadband operation, in a miniaturized (< 0.2 mm) configuration that can be integrated easily, and at frequencies > 2 THz, was not achieved to date, to the best of our knowledge.

Here, we present a SLG-based THz modulator with a tunable-by-design optical bandwidth. The electrostatic gating of SLG is achieved by a grating coupler, used as a gate electrode, and covered by $HfO_2/AlO_x$. This is patterned on a polyimide layer that acts as a quarter wave resonance cavity together with an Au reflector underneath. We achieve large modulation depths (90%) and ≥ 20 kHz electrical bandwidth (BW), in 1.9-2.7 THz range. We then integrate our device on-chip, with miniaturized THz QCL sources to alter the oscillating phase of the reflected intracavity field.

QCLs can inherently operate as frequency combs (FC) both in the mid-infrared [49] and THz spectral domains [50-54] through four-wave-mixing (FWM), spontaneously arising in the laser cavity as a consequence of the strong third-order non-linear susceptibility of the gain medium [52]. However, at THz frequencies, this only happens spontaneously over a very restricted operational range (< 23%) [50-54], in which the group delay dispersion is compensated. Alleviating this, by engineering and compensating the cavity dispersion over the entire dynamic range, is an extremely demanding task, since semiconductor materials are highly dispersive at THz frequencies. Current methods to achieve this rely on Gires-Tournois interferometer (GTI) [55] schemes with a biased [56] or an AFM-modulated section [57], or on tightly-coupled passive external cavity architectures [58,59], allowing extremely limited power outputs [57] or a limited increase of dispersion compensated regime [58,59].

Here we demonstrate that integration of the SLG modulator with the THz QCL, results in phase-locking of the laser modes, in an operational regime in which it cannot occur spontaneously, due to the inherent intracavity group delay dispersion. The integrated system then behaves as a frequency comb, spontaneously, over 35% of the laser operational range, therefore opening intriguing perspectives for tunable high-resolution spectroscopy and quantum metrology.

**Results and Discussion**

**Modulator Concept**

The design of the modulator, operating in reflection mode, is schematically shown in Fig.1a. To fabricate the device, we first deposit a 14 $\mu$m thick polyimide layer as a dielectric spacer (permittivity $\varepsilon_{poly}$=3.5 in the 2 - 4 THz range) on a Au/SiO$_2$/Si substrate (see Methods). The polyimide thickness is chosen to match the $n\lambda/4$ waveguide mode with *n (=1)* integer number (quarter wave mode) at the central frequency ν = 2.85 THz of our multimode QCL. Then we prepare a SLG capacitor on the polyimide layer as the active component of the device, as shown in Fig.1a (see Methods). The metal grating with a period *p* couples with the quarter wave mode and defines the modulator resonant frequency, while acting as the counter electrode of the SLG capacitor (see Methods).

To understand the optical characteristics of our devices, we perform finite element simulations with COMSOL Multi-Physics, using the Wave Optics Module. We model all the metal layers as perfect electrical conductor boundaries, and use Floquet periodic boundary condition, to create the metal grating. The optical constants of polyimide (Kapton) in the THz range are taken from Ref.60. We use the $\varepsilon_1 = 20$, thickness $t_1 = 40$ nm for $HfO_2$ and $\varepsilon_2 = 9$, thickness $t_2 = 100$ nm for $AlO_x$. We model SLG as a transient boundary condition with the Drude-like complex conductivity and with a Fermi energy dependent scattering time. We then calculate the spatial electric field (*E-field*) distribution of incident THz light with a polarization perpendicular to the grating at 2.48 THz. Fig.1b shows that *E-field* intensity is mostly focused to the grating edges. We also calculate the *E-field* distribution in the incident light direction, Fig.1c, for SLG on polyimide and Au reflector at 3 THz and for grating gated SLG at 2.48 THz, i.e. at the related λ/4 mode resonance frequencies. The intensity becomes zero on the Au reflector because of the discontinuity of *E-field* on a perfect conductor, and becomes maximum on the grating and SLG surfaces because of the quarter wave resonance. To simulate the reflected THz intensity modulation with the SLG $E_F$, we first calculate the fringing direct current (DC) electric field from the grating gate electrode using the AC/DC module in COMSOL (see Supporting information). The fringing field extends up to ~ 2.5 $\mu$m at the edge of the grating gate as a consequence of the finite impedance of the gate dielectric. We then use the calculated field profile to shift $E_F$ of SLG on top and on each side of the grating gate. We then simulate the reflectance of the grating gated SLG THz modulator with $p = 20$ $\mu$m, Fig.1d and for a set of different grating periodicities (Fig.1e).

The reflected light intensity decreases as $E_F$ increases, because of the increasing intraband absorption of SLG [17,20,37]. Then we consider the reflectance for different grating periods after setting $|E_F|$ ~ 200meV, in agreement with what retrieved from Raman spectra (see Supporting information). The quarter wave mode of SLG couples with the grating mode and shifts towards lower frequencies as the grating pitch increases from 5 to 40 $\mu$m, Fig.1e. By choosing the pitch size, we set the modulator spectral band in the 2-4 THz range, simultaneously keeping the grating mode outside this range as seen by the narrow resonance appearing at frequency 4.53 THz for a grating period $p = 40$ μm (green curve in Fig. 1e), corresponding to the 1st-order photonic mode of the grating gate cavity. This ensures a narrower band and a higher spectral tunability, when compared to the quarter wave resonance mode of SLG, whose linewidth is set by the polyimide thickness. A strong (> 50%) reflectance modulation of incident THz light can be achieved by tuning the $E_F$ by electrostatic gating after engineering the grating gate to move the quarter wave resonance to the targeted frequency.

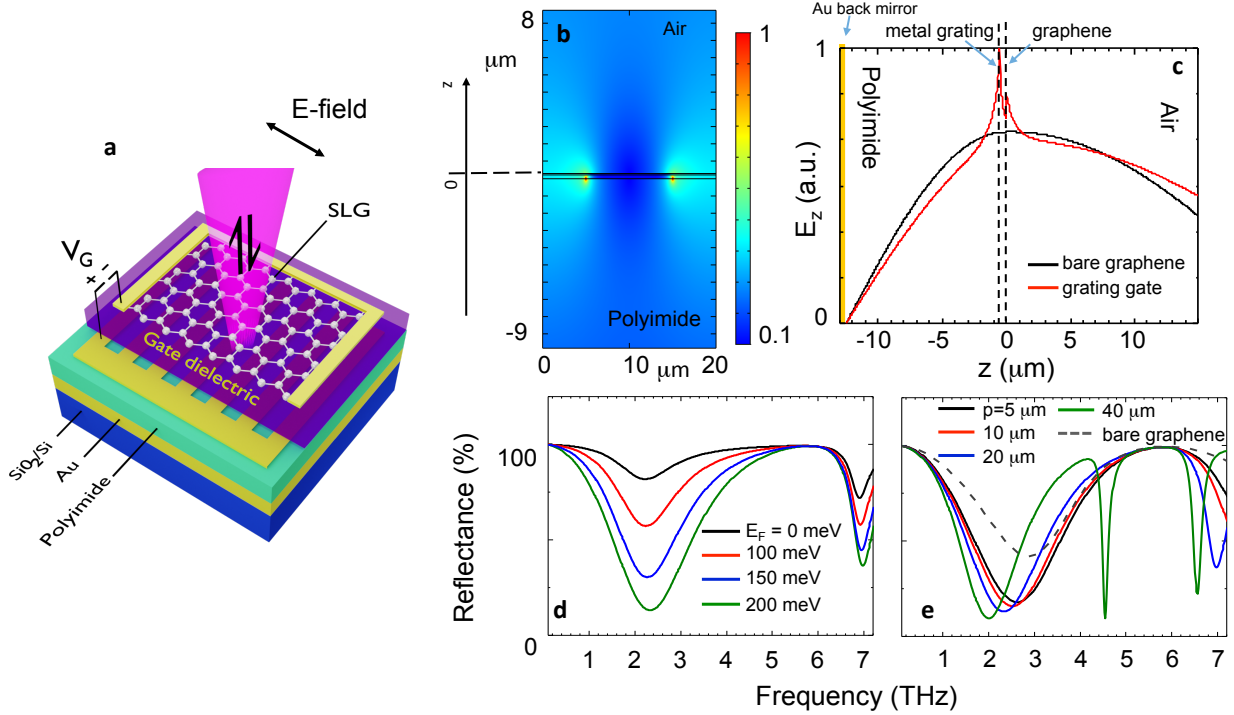

**Figure 1. Grating gated SLG THz modulator design**. **(a)** Schematic of modulator layout under top illumination, comprising a metal grating-SLG capacitor patterned on polyimide on Au. **(b)** Spatial distribution of the optical electric field on the metal grating-SLG capacitor, calculated for $p=20$ $\mu$m (metal width 10 $\mu$m) at the quarter wave resonance at $f = 2.48$ THz. **(c)** Electric-field intensity profile along the direction of incidence for SLG (black) at 3THz and for grating gated SLG (red) at 2.48 THz on polyimide. **(d)** Calculated reflectance of grating gated SLG modulator for different SLG $E_F$ for $p=20$ $\mu$m. To calculate the modulator reflectance at $E_F = 0$ meV, the graphene conductivity was set to the universal conductivity value $\sigma_0 = 6.08\times10^{-5}$ S [26]. The grating mode is ~ 6.88 THz. **(e)** Calculated reflectance for SLG (dashed line) and grating gated SLG THz modulators with different $p$ (continuous lines).

**Performance of the graphene THz modulators**

We measure the performance of our grating gated SLG modulators using a THz time domain spectroscopy (TDS) system (Menlo Terasmart k5) in reflection (see Methods and Supporting information). The time-domain THz signal is acquired with a delayed-pulse sampling window of 70 ps, resulting in a spectral resolution ~ 15 GHz. The beam spot diameter in reflection is 1.5 mm, i.e. smaller than the active area of the modulators (~2.5 × 4 mm$^2$). We measure the reflectance of SLG modulators with $p$ ranging from 5 to 45 $\mu$m at different gate voltages

.

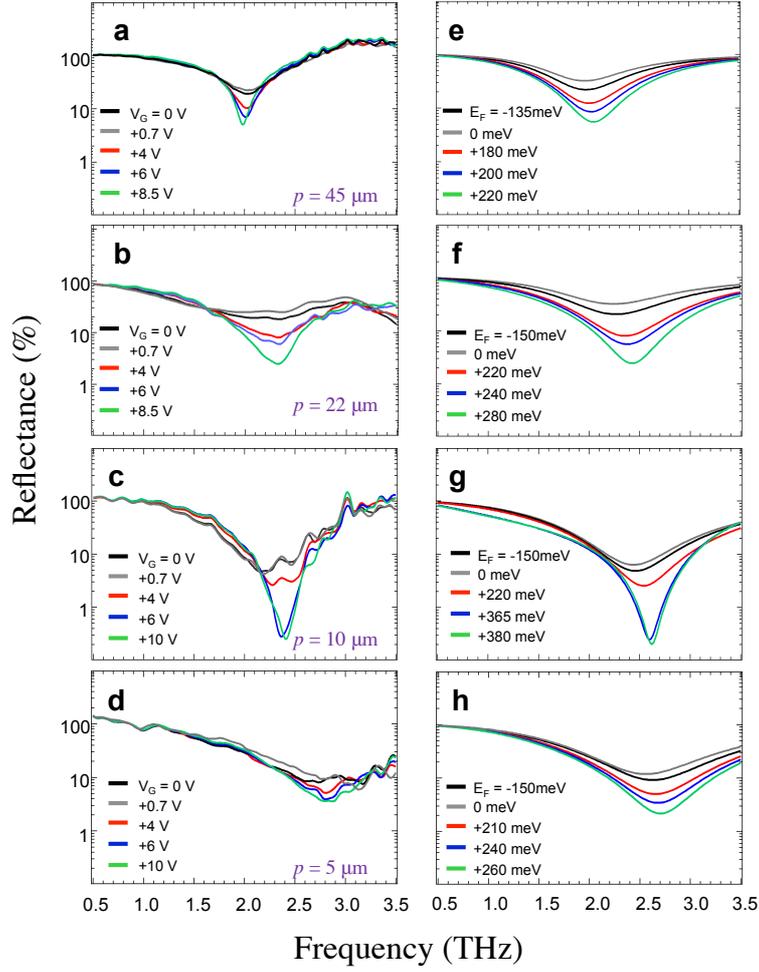

**Figure 2. TDS Reflectance spectra: (a-d)**. Total reflectance as a function of $V_G$ using a time-domain spectroscopy system for four modulators having different $p$ = (a) 45$\mu$m, (b) 22$\mu$m, (c) 10 $\mu$m, (d) 5 $\mu$m; **(e-h)** Calculated total reflectance for the same p of panels (a-d) at different $E_F$. In the simulated curves, the $E_F$ chosen to reproduce the experimental data are in agreement with a conductivity tuning corresponding to the experimental $V_G$, by assuming $n_0$ ~ 1.45 x $10^{12}$ cm$^{-2}$ for $p$ = 22, 45μm, and $n_0$ ~ 3.70 x $10^{12}$ cm$^{-2}$ for $p$ = 10, and $n_0$ ~ 3.35 x $10^{12}$ cm$^{-2}$ for $p$ =5 μm. The sign of $E_F$ is conventionally negative for p-type doping ($V_G<V_{CNP}$), and positive for n-type doping ($V_G>V_{CNP}$). In our sample, $V_{CNP}$ = +0.7 V as revealed by the IV measurements of the SLG FET (see Supporting Information), and the SLG was initially p-doped with a negative $|E_F|$ =135-150 meV.

The time-domain traces (Figs 2a-2d) reveal that the quarter wave resonance shifts from ~1.9 to ~2.9 THz as $p$ changes from 45 $\mu$m to 5 $\mu$m with the same thickness of polyimide (~14 $\mu$m). The reflectance from the modulators slightly increases as the gate voltage, $V_G$, changes from 0 to $V_{CNP}$ ~ + 0.7 V; then the reflectance decreases at $V_G > V_{CNP}$, due to the increasing charge density on SLG. $V_{CNP}$ is defined as $V_G$ at the charge neutrality point (CNP), (see Supporting information). $V_G$ is applied to the grating electrode. A $V_G$ value > $V_{CNP}$ shifts $E_F$ in the CB upward and accumulates electrons on SLG. The intraband transition rate gets higher with larger charge density, and the quarter wave resonance becomes deeper, due to the increasing intraband absorption of SLG [17,19,37].

We then simulate the optical performance of our SLG modulators using COMSOL Multi-Physics (see Methods and Supporting information) for the same device configurations as those experimentally measured. We enter SLG $E_F$, mobility and residual charge density, $n_0$, as variables in its Drude-like intraband conductivity to reproduce the experimentally measured reflectance. The $E_F$ dependent scattering time, $\tau = \mu E_F / e\, v_F$, where μ is the mobility, $e$ is the electron charge, $v_F$ is the Fermi velocity is then calculated by setting μ ~ 1600 cm$^2$V$^{-1}$s$^{-1}$ as measured from the electron transport characteristics of a field effect transistor (see Supporting information).

The calculated reflectances (Fig. 1e, Figs.2e-2h) are in broad agreement with experiments (Figs 2a-2d). The $E_F$ values in Figs. 2e-2h are compatible with those expected from the SLG electrostatic gating, assuming $n_0$ ~ 1.40 – 3.70 × 10$^{12}$ cm$^{-2}$ and of a gate capacitance $C_G$ = 65 nF/cm$^2$ (see Supporting Information). The corresponding conductivities match the values expected from the electrostatic gating of a graphene FET having the same gate capacitance of our modulator. The initial doping is that directly measured by Raman spectroscopy and the GFET IV characteristics. This $n_0$, due to the background free carrier density, gives a gate-independent contribution to the total reflectance at $V_G = V_{CNP}$. The highest doping concentrations are for modulators with the smaller grating pitch (p = 5,10μm), where the density of the grating edges is higher. These may increase the doping due to charge transfer [61]"

Although we can reproduce the modulation of the reflected THz light from each device, the expected crossing of the Dirac point and increase of the reflectance for negative $V_G$ is not seen. We attribute this to doping-inhomogeneity over the large 2.5 × 4 mm$^2$ area of the modulator, which prevents the homogeneus p-type carriers accumulation below the CNP.

To evaluate the modulation performances, we extract the modulation depth defined as: $\eta = 100 \times \left( \frac{R(V_G) - R(V_{CNP})}{R(V_{CNP})} \right)$ where $R(V_G)$ is the reflectance extracted by the TDS data at $V_G$, and $R(V_{CNP})$ is the maximum reflectance, measured at $V_G = V_{CNP}$. This figure of merit is a close estimate of the standard modulation depth, accounting for the reflectance variation relative to the Dirac voltage. For $p = 10\,\mu$m, and $V_G = +10$ V, $\eta$ reaches a maximum of 90% (Fig.3a). The corresponding insertion losses, defined as the ratio between the power reflected form the modulator and that reflected from an ideal Au mirror (R ~ 100%) calculated over the modulator optical bandwidth, are 1.3 dB, in agreement with the expected THz optical losses of SLG for $E_F$ ~ -200 meV (see Supporting Information).

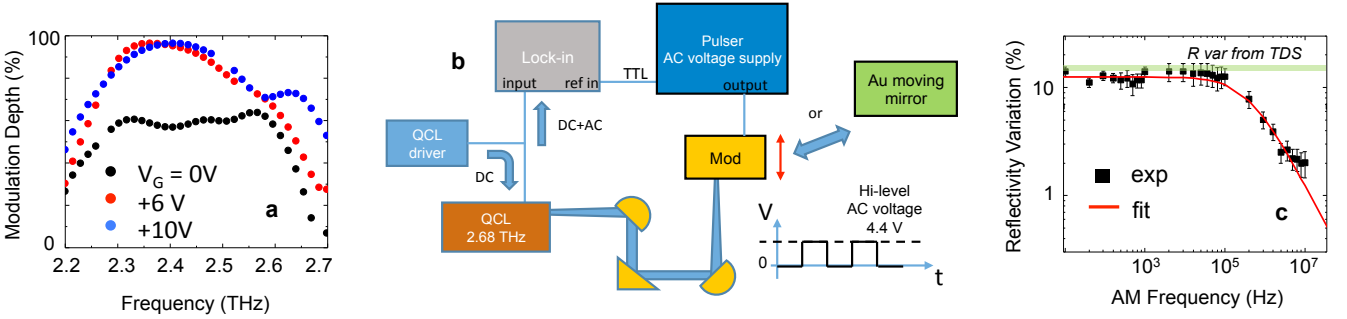

**Figure 3. Modulator depth and speed.** (a) Modulator depth calculated for $p$ =10 $\mu$m and resonance frequency 2.45 THz, with $V_G$ = -10V. (b) Schematic set-up to measure the modulation speed. The modulator is at the focal point of an external cavity comprising a single-plasmon CW QCL with single mode emission at 2.68 THz. The SM signal as a function of the AM frequency of the signal driving the modulator is recorded using a lock-in amplifier. The same configuration is used to acquire the reference SM signal, employing a moving Au mirror. The Au mirror is mounted on a AC-voltage-controlled piezoelectric stage, moving along the direction indicated by the red arrow. (c) Modulation speed, estimated through the reflectance variation ($\Delta R$), extracted from the QCL self-mixing signal as a function of the driving amplitude modulation speed in the set-up shown in (b).

To measure the modulation speed (MS), the modulator is placed at the focal point of an external cavity comprising a single-plasmon QCL which operates in continuous wave, CW, with single mode emission at 2.68THz. The single mode signal as a function of the amplitude modulation (AM) frequency driving the modulator is recorded by a lock-in amplifier. The same configuration is used to acquire the reference MS, employing a moving Au mirror, mounted on an AC-voltage-controlled piezoelectric stage moving along the direction indicated by the arrow in Fig.3b.

The modulation speed (Fig.3c) is then assessed through a detectorless approach exploiting self-mixing interferometry [62-64], i.e. by measuring the voltage change across the electrical contacts of a QCL operating at 2.7 THz, by employing the arrangement in Fig.3b. The physical principle is the intracavity reinjection of a small fraction ($10^{-4}$ - $10^{-2}$) of the emitted field that coherently interferes within the QCL cavity. In Fig.3b, the modulator acts as back mirror of an external cavity QCL. We therefore retrieve the down-converted signal arising from the electric field back-reflected by the modulator, that behaves as a gate-controlled active mirror of the external cavity QCL. The self-mixing trace then captures the reflectance variation corresponding to the applied AC gate voltage, hence the maximum modulation speed achievable while changing the amplitude-modulation (AM) frequency of the AC voltage.

We extrapolate the reflectance variation corresponding to the high- (+4.4 V) and low- (0 V) levels of the applied AC gate voltage from the TDS data. At the operating frequency of the QCL, such variation is ~ 15%, very close to the flat-region value of the measured self-mixing signal at low (< $10^3$ Hz) AM frequencies (Fig.3c). The 3dB cutoff frequency extrapolated from the fit, $f_{c.o.fit.}$=19.5 ± 1.8 kHz, is in very good agreement with the theoretical electronic cutoff $f_{c.o.teo}$ = 24.5 kHz, obtained approximating the modulator with an equivalent low-pass RC circuit having R= 1 kΩ (the SLG resistance, measured directly on the sample with a two-probe method) and C = 6.5 nF, the overall modulator capacitance. The latter is calculated as C = $C_G$ × $A_{mod}$, where $A_{mod}$ = 10 mm$^2$ is the total modulator area, and $C_G$ is obtained as the series of the capacitances of the two dielectric layers: HfO$_2$ ($\varepsilon_1$ = 20, thickness $t_1$ = 40 nm, $C_1$ = 450 nF/cm$^2$), and AlO$_x$ ($\varepsilon_2$ = 9, $t_2$ = 100 nm, $C_2$ = 76 nF/cm$^2$). The measured speeds (50 μs) arise from a combination of device size and optimization of optical coupling. Thus, our design has the potential to achieve MHz modulation bandwidths, by combining device scaling and optical coupling optimization. E.g., by reducing the modulator area to that of the facet of a THz QCL micro-laser (15μm × 85μm), the parasitic capacitance would decrease by~4 orders of magnitude, meaning that, even if the graphene resistance is partially affected by the more rectangular geometry, the modulator electrical bandwidth could reach 40MHz"

**Integration of SLG-THz modulator with a QCL-GTI**

We then investigate the performance enhancement induced by our SLG modulators on the intracavity mode dynamics of a heterogeneous multimode THz QCL, by coupling a SLG modulator with a QCL. The QCL comprises a 17-μm-thick GaAs/AlGaAs heterogeneous heterostructure, featuring a sequence of three active region modules with frequency-detuned gain bandwidths [54] centered at 2.5 THz, 3.0 THz, and 3.5 THz [65]. The Fabry-Perot device operates as FC synthesizer over ~ 1.05 THz bandwidth and shows a stable and narrow (4.15 kHz) beatnote over a continuous current range ~ 106 mA, corresponding to 15% of the laser operational range. Differently to previous approaches [66], we then engineer an etalon-like interferometric scheme, in which the modulator is tightly coupled to the back facet of the THz QCL, at a distance~50μm, required for an on-resonance GTI [58]. The THz radiation is injected in the laser cavity after being back-reflected by the modulator, operating as the external cavity active mirror of a GTI. The motivation is to exploit the comb sensitivity to the optical feedback in the modulator-GTI configuration, to achieve partial compensation of the group delay dispersion (GDD) in the QCL cavity. The shape of the modulator reflectance, and its resonant architecture, in the explored frequency range, can help minimize the total cavity dispersion, since the phase of the reflected light and, therefore, the GDD oscillates according to the pre-defined modulator resonance. To verify this, we calculate the GDD for the QCL-modulator coupled system for $p$ = 5 μm, 10 μm, and 22 μm, chosen to cover 3 distinctive configurations: (i) when the modulator optical bandwidth is superimposed perfectly with the QCL gain bandwidth ($p = 5\mu$m); (ii) when it is shifted towards the low-frequency side of the laser bandwidth ($p = 10\ \mu$m); and, (iii) when it is completely detuned ($p = 22\mu$m).

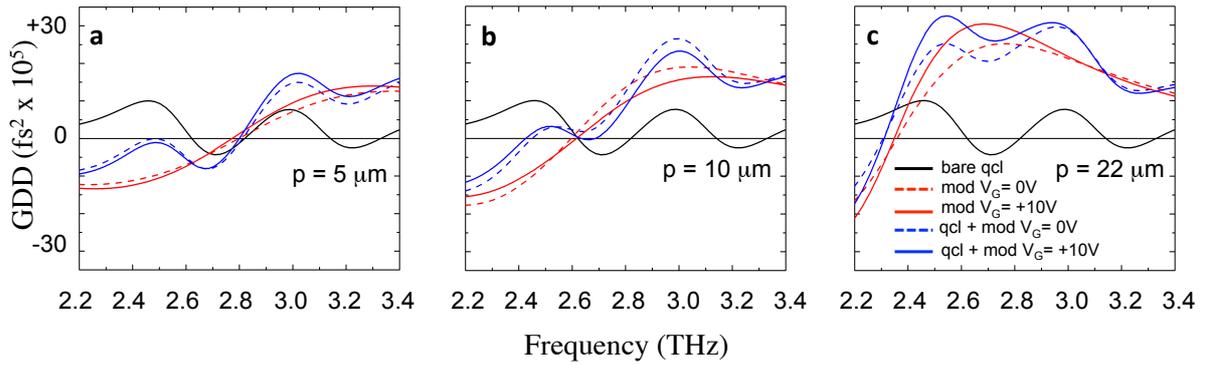

**Figure 4. GDD of QCL-modulator integrated system: (a-c)** GDD of the heterogeneous QCL frequency comb calculated as a function of frequency when the QCL is uncoupled (black) or coupled with the SLG modulator (blue), and of standalone SLG modulator (red), at $V_G$ = 0V (continuous line) and $V_G$ = +10V (dashed line), for (a) $p$=5$\mu$m; (b) $p$=10 $\mu$m; and, (c) $p$=22 $\mu$m.

The calculated GDD curves in Figs.4 are obtained from the sum of the individual contributions arising from the GTI-QCL and from the modulator only (see Methods). The anti-symmetric profile of the modulator GDD comprises a low frequency negative region, which changes sign towards higher frequencies; the zero-crossing point corresponds to the measured absorption resonance. The GDD amplitude increases while varying $E_F$. In the coupled system, the total GDD is reduced in the low-frequency region of the QCL gain bandwidth (~2.2-2.7 THz), particularly for $p = 5\ \mu$m (Fig.4(a)) and $p =10\ \mu$m (Fig.4(b)).

Furthermore, in the investigated $p$ range, the integration of the modulator is expected to have a detrimental effect at frequencies > 2.7 THz, especially for larger $p = 22\ \mu$m (Fig.4(c)) and mismatched

QCL/modulator bandwidths. As a general trend, the GDD amplitude increases for larger *p* due to the higher optical strength and lower optical losses of the *nλ*/4 grating-gate waveguide mode, as reflected in the stronger absorption mimima in the TDS plots of Figs. 2.

Consistently, a similar behavior is also predicted for increasing $E_F$. The simulations indicate a weak gate-tuning of the total GDD, for the same $E_F$ (100 - 300 meV) used to describe the gate modulation of the reflectance (Figs. 2). As a consequence of the higher modulator GDD, the compensation of the total GDD over the whole QCL band becomes less effective.

To explore the efficacy of the coupled cavity configuration in modifying the intracavity QCL mode dynamics, we trace the QCL intermode beatnote map by employing the set-up in Fig. 5a.

The comparison between the CW emission spectra of the bare laser (Fig. 5b) and the modulator-coupled QCL (Fig. 5c) does not show significant differences in mode distribution, spacing, spectral coverage and relative intensities.

The electrical beatnote maps for 3 SLG modulators with different p (Figs.5d-f), show changes in the intracavity dynamic range, when compared with the corresponding maps measured on the bare QCL [54] (Supporting Information, Fig. S8), or with that collected coupling the same laser with an Au mirror placed at the same distance [58]. Furthermore, the different *p* and related reflectance spectra affect the QCL intracavity mode dynamics. Coupling the QCL with SLG modulators with *p* = 5, 10 *µ*m (Figs 5d,e), we first observe a region of single and narrow (3.8 kHz, Fig. 6a) beat-note extending over a continuous current range of 165 mA (440 – 605 mA), significantly larger than that of the corresponding bare QCL (106 mA) [54]. At larger currents (605– 760 mA) we observe, in both cases (Fig. 5d-5e), a broad beatnote characteristic of a lasing regime in which the group velocity dispersion is large enough to prevent locking of the lasing modes in frequency and phase, simultaneously, via four-wave mixing.

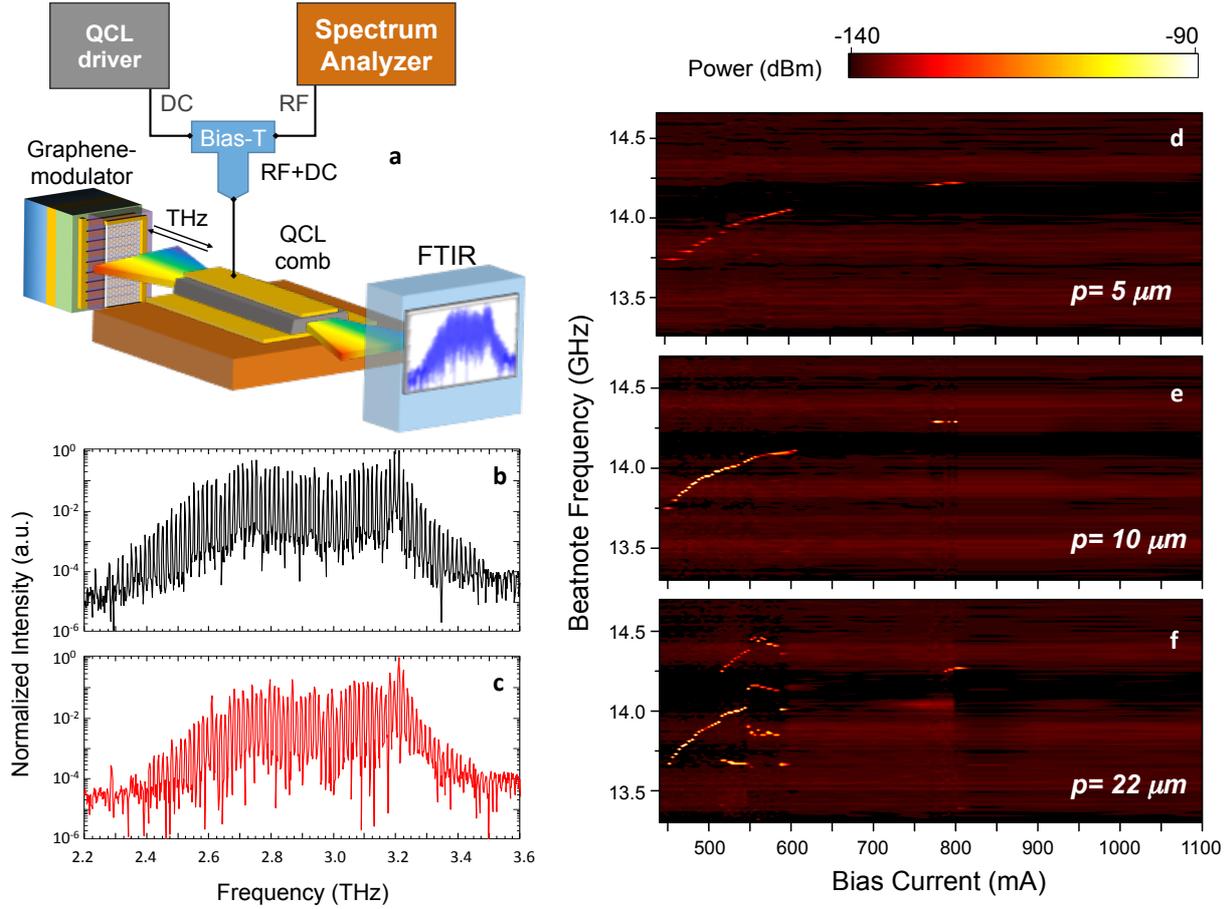

**Figure 5. QCL frequency comb embedding a SLG modulator:** **(a)** Schematic experimental setup. The SLG modulator is positioned on a moving piezoelectric stage in close proximity (50μm) to the back facet of a multimode heterogeneous THz QCL. The modulator is coupled on the same copper mount with the QCL onto the cold-head of a He-flow cryostat. In this configuration, the back-reflected radiation is injected into the QCL waveguide, while the emission from the front facet is collected into a Fourier-Transform InfraRed (FTIR) spectrometer. The intermode beatnote maps are recorded with an RF spectrum analyzer (Rohde & Schwarz FSW). **(b-c)** FTIR emission spectra at 15K, under vacuum, with~0.075cm$^{-1}$ resolution, whilst driving the QCL in CW with 780 mA from (b) bare QCL and (c) QCL-modulator tightly coupled system, for $p = 10$ $\mu$m at $V_G$= 0V. **(d-f)** Intermode beatnote maps as a function of CW driving current at 15 K in the QCL-modulator system with (d) $p = 5\mu$m, (e) $p = 10$ $\mu$m, (f) $p = 22$ $\mu$m. The beatnote signal is extracted from the bias line using a bias-tee with a RF spectrum analyzer, and is recorded with resolution bandwidth (RBW): 500 Hz, video bandwidth (VBW): 500 Hz, sweep time (SWT): 20 ms, RMS (root mean square) acquisition mode.

On the other hand, when coupling the QCL with a $p = 22$μm modulator, a single and narrow beatnote appears only in a very restricted operational range (440 -520 mA), roughly comparable with that observed in the bare QCL [54]. Very differently from the bare QCL [54], at higher currents (520-600 mA), the device shows a region of three narrow beatnotes. In the bare laser, we first observe multiple beatnotes, indicating lasing from higher order lateral modes, then a single beatnote and finally double beatnotes, reflecting the dual comb behavior of the laser [54]. Conversely, in Fig. 5f, multiple beatnotes persist from 520 to 600 mA unveiling a complex intracavity dynamics [67] and the lack of GDD compensation, as predicted by simulations (Fig. 4c). The tuning coefficients of the 3 beatnotes in the 515-558 mA and 558-600 mA ranges are dissimilar, indicating that, while in the first portion of the dynamic range the 3 active regions embedded in the QCL core are individually behaving like FCs, in the second current range, higher order lateral modes can interfere with the main QCL modes, preventing comb operation [68].

In all cases (Fig.5d-f) the beatnote turns again single and narrow (7.6 kHz, Fig.6b) at larger driving currents in the range 760 – 820 mA, while coupling the QCL with the SLG modulators having $p = 5\mu$m and $p = 10\ \mu$m, and over a slightly narrower range (780 – 820 mA) when $p = 22\ \mu$m. In this new operating regime, not observed either in the bare-QCL FC [54], or while coupling the same QCL with an Au mirror [58], we retrieve a stable, high-intensity (30dBm), narrow (5-7 kHz) single beatnote with an emission spectrum covering a 1.2 THz bandwidth (Fig. 5c), with 98 equidistant optical modes. The coupling with the SLG modulators results in a stable frequency comb operation over 35% of the laser operational range, as extracted from the data of Fig. 5e and 6e, much larger than that reported in any previous passive THz QCL frequency comb [53, 54], to the best of our knowledge. Such an effect is almost independent of $V_G$, as can be seen in plot of the intermode beatnote linewidths for different $V_G$ in the modulator with $p=10\mu$m (see Supporting Information). The beatnote map and the QCL emission spectra remain almost the same.

The evolution of the emission spectra measured in the bare heterogeneous QCL [54] shows that the 3 active regions each possesses slightly different threshold current densities [65], and that the active region reaching threshold at larger bias is the one centered at the lowest frequency (2.5 THz). This matches the spectral region in which the modulator-QCL GDD compensation is more effective (Figs 4a-4c). Therefore, the appearance of the single beatnote regime at the higher bias is likely to be result of dispersion compensation, simultaneously preventing higher-order lateral modes reaching threshold, and locking in phase the family of modes arising from the active region module that is centered at lower frequency (2.5 THz).

We can exclude that phase chirping effects upon modulation can induce a relevant effect on the cavity dispersion of our heterogeneous THz QCL, since the modulator is not inducing any visible and detectable phase modulation in the QCL.

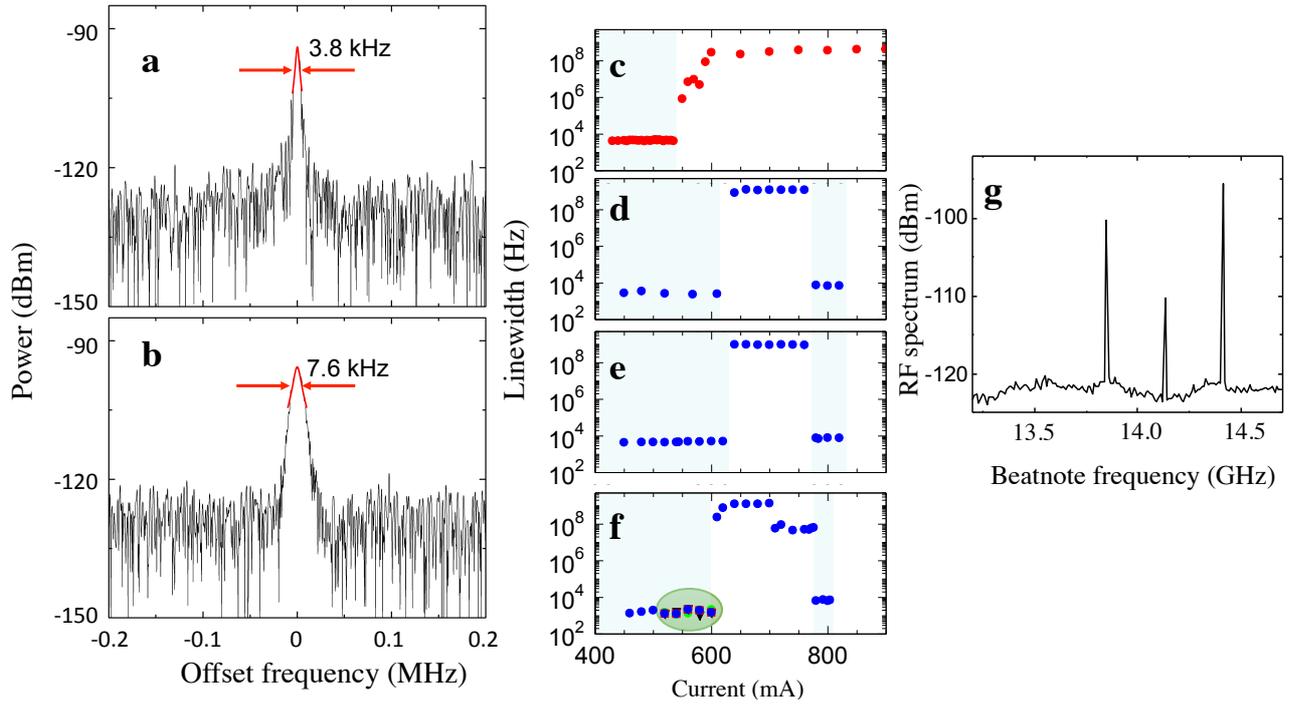

**Figure 6. Analysis of the intermode beatnote linewidths. (a-b)** Intermode beatnote signal from the QCL integrated with the SLG modulator with $p =10$ $\mu$m at (a) 500, (b) 800mA, measured in CW, at 15 K. The RF spectrum analyzer settings are: RWB: 10 kHz, VBW: 100 kHz, SWT: 200 ms, RMS acquisition mode. **(c-f)** Evolution of the intermode beatnote linewidths as a function of QCL driving current for (c) bare-cavity QCL[46], or when the QCL is coupled with the SLG modulators with (d) $p = 5$ $\mu$m; (e) $p = 10$ $\mu$m; (f) $p = 22$ $\mu$m. The red and green marks in (f), identified via the green shaded ellipse, correspond to the linewidths of the multiple beatnotes shown in (g). **(g)** Intermode beatnote signal acquired from the QCL integrated with SLG modulator with $p =22$ $\mu$m at 560mA. The light blue shaded areas identify the region in which the laser behaves as a stable frequency comb synthesizer.

The analysis of the intermode beatnote linewidth (Figs. 6c-f) shows that at driving currents ≤ 605 mA, the intermode beatnotes linewidth values are comparable (3 – 5 kHz, Fig.6d-f) with those measured on the bare laser (Fig. 6c), and become slightly narrower (1.5 kHz) when the QCL is coupled with the $p = 22$ μm modulator. Such a linewidth narrowing applies to each individual multiple beatnote (Fig. 6g), as an effect of their individual behavior as homogenous QCL frequency comb [68].

In the 780 – 820 mA range, the beatnote linewidths becomes larger and range between 6 kHz and 8 kHz, as an effect of the jittering induced by the partial increase of the lattice temperature during CW operation [54].

In the 605 mA – 780 mA range, as in the case of the bare-laser cavity, we retrieve the typical very high phase noise (beatnote linewidths ~$10^8$-$10^9$ Hz ranges, the broad band retrieved on the spectrum analyzer with a Lorentizian [50,51,53,54,56,58]) resulting from the GDD-induced entanglement of dispersion.

**Conclusions**

We demonstrated a modulator design, comprising a grating-gated graphene capacitor on a polyimide quarter wave resonant cavity. It ensures frequency scaling by-design and efficient (90%) amplitude modulation at high speeds (> 20 kHz). It can be extended to alternative photonic concepts, such as spatial light modulators or nano switches. The ease of implementation and flexible design can allow integration with existing laser sources to alter their intra-cavity field and mode dynamics. The stable frequency comb operation, close to the laser peak power, and over a dynamic range (35% of the laser operational regime) much wider than other passive approaches [42-46], proves the versatility of our modulator. Our integrated graphene modulator-QCL has a great potential for high-resolution and high-precision spectroscopy [68], and quantum metrology [68,70]. It could also be integrated in a compact, Peltier-cooled portable configuration, that can operate at 250K [71]. A miniaturized frequency comb delivering > 1 mW optical powers, with > 90 equidistant modes, tunable over 35% of the operational range of the laser can allow applications never addressed so far, such as manipulation of cold atoms and molecules, sensing in space science, and entanglement of the QCL optical modes, crucial for one-way quantum computing.

**Methods**

**Device Fabrication.**

To fabricate the grating-gated SLG THz modulators, we first spin-coat a 14-$\mu$m thick polyimide layer on a Au(300nm)/SiO$_2$(350nm)/Si back-mirror and bake it at 350 °C for 30 minutes. We then prepare a metal grating on polyimide layer and deposit 100 nm AlO$_x$ by sputtering and 40 nm HfO$_2$ by atomic layer deposition (ALD).

SLG is then grown in a hot wall chemical vapor deposition (CVD) system on Cu foil (35$\mu$m, Graphene Platform, Japan). The foil is loaded into a quartz chamber in a horizontal furnace and the system evacuated to ~10 mTorr base pressure. 40 sccm H$_2$ is then introduced into the system to attain 400 mTorr pressure and the growth chamber is heated to 1000ºC. The Cu foil is annealed in the same conditions for 30 min. To initiate the growth of SLG, 5 sccm CH$_4$ is introduced into the system and the growth is terminated after 30 min by turning off all the gases and the heater. The system is then naturally cooled down to room temperature. SLG is then transferred on AlOx/HfO$_2$ using wet transfer [31, 36]. Polymethyl methacrylate (PMMA, A4 950, MicroChem) is spin-coated onto the SLG/Cu foil at 3000 rpm for 60s and baked at 120ºC for 5 min. The PMMA/SLG/Cu stack is then floated on a solution of 0.1M Ammonium persulfate (APS) overnight to etch the Cu foil and then moved in DI water to clean APS residues. The floating PMMA/SLG stack is then transferred on AlO$_X$/HfO$_2$/polyimide/Au/SiO$_2$/Si, Fig.1a. The sample is then dried in ambient for 2 hours and subsequently baked at 120 ºC for 15 min. PMMA is then removed in acetone and IPA. 5/100nm Cr/Au are deposited by thermal evaporation for both metal grating gate and the contact. Metal gratings with $p$ = 5, 10, 22 and 45 µm are fabricated with a 50% geometrical fill factor (i.e. the metal width equals the metal gap in the unit cell). As grown and transferred SLG samples are characterized by Raman spectroscopy using Renishaw InVia spectrometer (Supporting Information).

**Full-wave electromagnetic simulations**

The complex reflection coefficient and total field distribution under single-frequency illumination of the devices are simulated using the finite element method software COMSOL Multiphysics in the frequency domain. SLG is simulated as a monolayer (0.34 nm) with intraband Drude-like conductivity. The polyimide's (Kapton) optical constants in the THz range are taken from Ref. 60, while Au is modeled as a perfect electric conductor. The structures are illuminated with single frequency light using a port boundary condition, and the amplitude and phase of the reflection coefficient are extracted from the simulation. The resulting GDD is calculated via the second derivative of the phase of the reflection coefficient. The GDD of the bare laser is calculated as in Ref. 58.

**Modulation speed.**

We measure the modulation speed from a self-mixing (SM) experiment, Fig. 3c. The self-mixing signal is the down-converted signal associated with the cavity optical feedback experienced by the QCL, once positioned inside an external cavity, comprising the modulator placed at one vertex of the QCL beam path. The modulator is driven by a voltage oscillating at the AM frequency generated by a pulsed driver (Agilent 8114A). The self-mixing signal is then recorded using a lock-in amplifier (Zurich Instruments UHFLI). A single-plasmon CW QCL and emitting single-mode emission at 2.68 THz is then coupled with the $p$ = 10 $\mu$m modulator. The modulator driving gate AC voltage is a square-wave signal with amplitude 4.4 V, baseline 0 V, duty cycle 50% and AM frequency varying from $10^2$ to $10^6$ Hz. For an accurate estimate of the reflectance $R$, we extract the reference level by the self-mixing signal recorded with a moving Au mirror, oscillating at the same AM frequency. The reference SM signal is almost flat in the AM frequency range covered in the present experiment.

**Modulator-QCL integration.** The grating-gated SLG THz modulator is integrated with the THz QCL into an external etalon-like configuration, defining a GTI (Fig.5a). The QCL is a 2.9-mm-long, 85-µm-wide Fabry-Pérot laser bar operating as a frequency comb synthesizer in CW over 15% of its dynamic range. The QCL is mounted on a Cu bar in thermal contact with the cold-finger of liquid-He cryostat, with one of the two emitting facets positioned at the vertex of the input optical path of a vacuum Fourier-transform infrared (FTIR) spectrometer (Bruker vertex 80), for the simultaneous collection of the QCL emission spectra. FTIR spectra are collected in rapid-scan mode, under vacuum, with a 0.075 cm$^{-1}$ resolution, at 15 K. The SLG modulator is integrated with the QCL inside the cryostat with the help of a piezoelectric stage, enabling tight coupling of the modulator active area with the QCL facet, at a 50 μm distance. The intermode beatnote map is acquired as a function of laser drive current at a heat sink temperature of 15 K. The beat-note signal is extracted from the bias line using a bias-tee, and is recorded with a RF spectrum analyzer (Rohde & Schwarz FSW; resolution bandwidth (RBW): 500 Hz; video bandwidth (VBW): 500 Hz; sweep time (SWT): 20 ms; RMS acquisition mode).

**Acknowledgements**

We acknowledge funding from ERC SPRINT (681379), Hetero2D, GSYNCOR, EU Graphene Flagship (core III), EPSRC Programme Grant 'HyperTerahertz' ((EP/P021859/1), EP/K01711X/1, EP/K017144/1, EP/N010345/1, EP/L016087/1,the Royal Society and Wolfson Foundation